Title: 3-D off-center Li$^+$ rotors in alkali halides: solving for the eigenvalue problem
Author: Mladen Georgiev (Institute of Solid State Physics, Bulgarian Academy of
  Sciences, 1784 Sofia, Bulgaria)
Comments: 10 pdf pages with 6 figures
Subj-class: physics

We are dealing with the long-standing problem of the eigenstates and eigenvalues of 3-D rotators by off-center impurity ions in alkali halides. The ion runs along the brim of a sombrero-like vibronic potential. The quantum-mechanical motion depending on two angular coordinates, we consider the motion along θ at averaged φ, and vice versa. We find the off-center ion performing two rotations: one in (x,y) plane along the azimuth φ and another one along the position angle θ, respectively, which are both described as Mathieu's non-linear oscillators. The motion along the z-axis is complemented by a $z^4$ anharmonic mode.

1. Introduction

The off-center ions constitute one of the most thrilling discoveries in solid state physics [1]. As smaller-size isolated substitutional impurities are incorporated in an ionic crystal, they go off-center to give rise to a number of specific features, among them the appearance of inversion dipoles, due to the breaking up of site inversion symmetry, with associated paraelectric effects, a rotation-like reorientation motion with the appearance of magnetic dipoles, with associated paramagnetic effects, as well as the appearance of numerous fingerprints imposed on the optical spectra and the paraelectric behavior of the impurity [2], etc. According to a viewpoint widely shared now, the off-center displacements are due to the quantum-mechanical vibronic effect, whose generator is the mixing up of nondegenerate electronic states at the impurity by a coupled vibrational mode [3]. However, there are other viewpoints too, shared by a fewer audience, though, which attributes the displacements to classical polarization interactions [1] or to chemical rebonding [4]. We have so far found no evidence to contradict the quantum-mechanical interpretation but a few to render the Jahn-Teller effects more likely [5]. Depending on the electronic states and the coupled vibrations, whether local states or narrow bands, the vibronic effect (sometimes called pseudo-Jahn-Teller effect too) gives rise to local distortions [6] or to polaron ones when the distortion moves along with the charge carrier associated to it [7]. It appears that vibronic polarons may be behind a number of puzzling phenomena discovered lately, such as colossal magneto- and electro- resistances, to mention a few [8].

2. Hamiltonian

The coupling Hamiltonian is of the electron states mixing type (non-quantized lattice):

$H_{int} = \Sigma_{\alpha\beta} P(G_{\alpha\beta}Q_{\alpha\beta})[a_\alpha^\dagger a_\beta + h.c.]$ (1)

where $P(q_{\alpha\beta})$ is a polynomial of the vibrational coordinate $Q_{\alpha\beta}$ of the mixing vibrational mode, $G_{\alpha\beta}$ is the mixing constant, $a_\alpha$, etc. are the electron ladder operators. α, β are electronic states

(local defects) or narrow bands (vibronic polarons). By virtue of group theory, which rules that the representation of the point group for the vibrational mode (phonons) must be included in the direct product of the representations of the electronic states (bands): $\Gamma_Q \subset \Gamma_\alpha \otimes \Gamma_\beta$, and since $\alpha$ and $\beta$ are opposite-parity states (bands), $\Gamma_q$ must be odd-parity. For this reason, $P(Q_{\alpha\beta})$ is to be composed of odd powers of $Q_{\alpha\beta}$ only, the even powers leading to vanishing mixing constants, according to group theory [9].

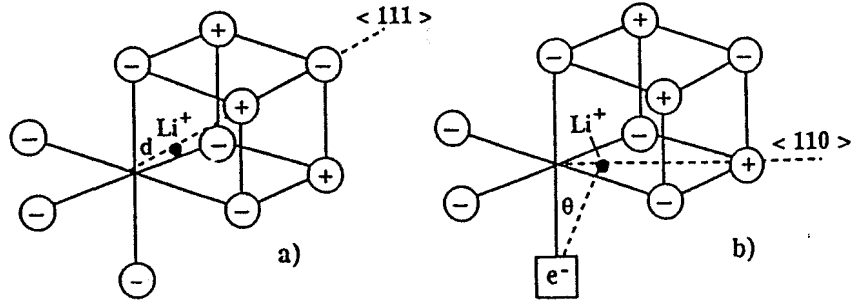

Figure 1(a),(b): Isolated Li+ impurity (a) and one nearest-neighboring an F center (b) [16].

The isolated Li$^+$ impurity has been found to occupy an off-site position along $\langle 111 \rangle$ in the cubic alkali halide host, as shown in Figure 1(a), while it occupies a $\langle 110 \rangle$ site if perturbed by a nearby F center at [001], as in Figure 1(b). In addition, the experiments have evidenced small off-axis displacements in both cases, as sketched on these same drawings. The origin of these off-axis displacements, as found in optical experiments, is not yet fully understood. From Reference [16].

It has been proven that on the above parity premises the off-center displacements are obliged to the 1$^{st}$ order coupling constant, while the 3$^{rd}$ order constants bring in new features [10]. To be exact, the 1$^{st}$ order creates smooth off-center rings or ellipses as rotation paths of the displaced impurity, while the 3$^{rd}$ order constant embroiders the rotation paths by barriers along with valleys corresponding to the orientation sites.

A complete account of our investigation of the eigenvalue problem with the coupling operator as in (1) has been described elsewhere [11]. We shall now concentrate on the adiabatic (vibronic) energy for the local polaron. Following it we obtain (up to constant terms)

$$E_{AD\pm}(Q_{\alpha\beta}) = \tfrac{1}{2} \{KQ_{\alpha\beta}^2 \pm \sqrt{[4P(G_{\alpha\beta}Q_{\alpha\beta})^2 + E_{\alpha\beta}^2]}\} \qquad (2)$$

where $E_{\alpha\beta}$ is the energy gap between the basis electron states. We remind that because of the group theory rule

$$P(G_{\alpha\beta}Q_{\alpha\beta}) = \Sigma G_{\alpha\beta j}Q_{\alpha\beta j} + \Sigma G_{\alpha\beta ijk}Q_{\alpha\beta i}Q_{\gamma\delta j}Q_{\lambda\mu k} \qquad (3)$$

which is reduced to (here and above site and band indexes are dropped for better clarity)

$$P(G_{\alpha\beta}Q_{\alpha\beta}) = G_{\alpha\beta}\Sigma Q_{\alpha\beta j} + G_{\alpha\beta iii} \Sigma Q_{\alpha\beta i}Q_{\gamma\delta i}Q_{\lambda\mu i} + G_{\alpha\beta iik}\Sigma Q_{\alpha\beta i}Q_{\gamma\delta i}Q_{\lambda\mu k} \equiv$$

$$G\Sigma Q_i + D_b\Sigma Q_iQ_iQ_i + D_c\Sigma Q_iQ_iQ_k \text{ for } D_c = G_{\alpha\beta iii}, D_b = G_{\alpha\beta jkk}, 0 = G_{\alpha\beta ijk}. \qquad (4)$$

Equation (4) comes to stress the importance of the diagonal elements over the rest in shaping the coupling to the $T_{1u}$ vibrational mode.

Equations (2) and (4) give the vibronic potential energy to be inserted into the Schrödinger equation to solve the vibronic problem. We get accordingly (the summation is over the Descartes components of the configuration coordinates:

$$-(\hbar^2/2M)\Delta_{\rho\theta\varphi}\,\chi + E_{AD\pm}(Q)\,\chi = E\,\chi$$

$$E_{AD\pm}(Q) = \tfrac{1}{2}\{KQ_{\alpha\beta}^2 \pm \sqrt{[4(G\Sigma Q_i + \Sigma D_{ijk}Q_iQ_jQ_k)^2 + E_{\alpha\beta}^2]}\}$$
(5)

We next perform the operations under the square root to obtain (neglecting all powers of Q higher than the fourth in view of the small-displacement coordinates):

$$E_{AD\pm}(Q) = \tfrac{1}{2}\{K\Sigma Q_i^2 \pm \sqrt{\{[(G\Sigma Q_i) + (\Sigma D_{ijk}Q_iQ_jQ_k)]^2 + \tfrac{1}{4}E_{\alpha\beta}^2\}}\}$$
(6)

and using the specific tensor model on introducing the radial coordinate $Q = \Sigma Q_i^2$:

$$E_{AD\pm}(Q) = \tfrac{1}{2}KQ^2 \pm \sqrt{\{(GQ)^2 + 2G[(D_c - D_b)\Sigma Q_i^4 + D_bQ^4] + \tfrac{1}{4}E_{\alpha\beta}^2\}}$$
(7)

$E_{AD\pm}(Q)$ is a multidimensional double-branch surface in Q-space. Its maximum dimension will be limited to 4. This suffices for predicting most of the essential features of the coupled system.

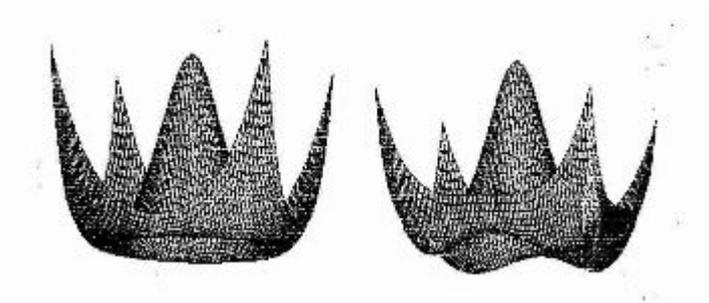

Figure 2(a),(b): Lower-branch vibronic potentials to 1$^{st}$ (a) & 3$^{rd}$ (b) order mixing [17].

At $D = 0$ we obtain the response of the coupled system to off-center displacements $Q_{0i}$. The latter result in a net displacement of the ion metastable position along an off-center surface (in 3-D) or ring (in 2-D), circular or elliptic, depending on the coupled mode symmetry. In the circular case the off-center displacement is obtained by minimizing $E_{AD-}(Q)$:

$$E_{AD\pm}(Q) = \tfrac{1}{2}KQ^2 \pm \sqrt{[(GQ)^2 + \tfrac{1}{4}E_{\alpha\beta}^2]}$$

Thus we get the extremal radius of the off-centered surface:

$$Q_0 \equiv \Sigma Q_{i0}^2 = \sqrt{(2E_{JT}/K)}\sqrt{[1 - (E_{\alpha\beta}/4E_{JT})^2]}$$
(8)

where $E_{JT} = G^2/2K$ is the Jahn-Teller energy. At the later stage we will introduce spherical coordinates $(Q_0,\theta,\varphi)$ to account for the extremal features of $E_{AD}(Q)$: It is a sombrero-shaped surface (in 2-D), as shown in Figure 2(a), with a central maximum at $Q = 0$ and a circular (or elliptic) valley (hat brim) at radial distance $Q_0$ from the origin in the basic (x,y) plane (normal lattice site). $E_{AD}(Q)$ is extremal along the brim setting the radial derivative $\partial/\partial r$ to vanish to a reasonable extent all along in the brim valley. Also vanishing along a circular ring are the angular derivatives $\partial/\partial\theta$ and $\partial/\partial\varphi$ because of the spherical symmetry. As a result, the Laplace operator in spherical coordinates $Q_x = Q_0 \sin\theta \cos\varphi$, $Q_y = Q_0 \sin\theta \sin\varphi$, $Q_z = Q_0 \cos\theta$ (x,y,z = i,j,k) along the hat brim:

$$\Delta_{Q,\theta,\varphi} = (r^2\sin\theta)^{-1}[(\partial/\partial r)[(r^2\sin\theta)(\partial/\partial r)] + (\partial/\partial\theta)[(\sin\theta)(\partial/\partial\theta)] + (\partial/\partial\varphi)[(1/\sin\theta)(\partial/\partial\varphi)] \quad (9)$$

is mainly controlled by $\Delta_Q = \partial^2/\partial Q^2$.

At $D \neq 0$, the hat brim becomes embroidered by rotational barriers, each barrier lying between two neighboring orientational axes at the off-center sites around the normal lattice site. The rotational barriers hinder the nearly-free rotation along the brim. These barriers are split by rotational valleys which hold the metastable orientation sites of an off-center ion. The reorientation barriers are shown in Figure 2(b) for a 2-D system. From Reference [17].

At $D = 0$ the off-center ion performs nearly-free smooth rotations along the sombrero-hat brim which turn barrier controlled at $D \neq 0$. The in-plane rotation reminding of the equatorial rotation of a spinning particle, it can be ascribed a chirality, as in Figure 3. However, this feature of the reorientating off-center ion will be an *orbital chirality* unlike the *spin chirality* of the neutrino. The chirality feature will be needed a due attention in a future study. From Reference [18].

Two other Li$^+$ configurations in the relaxed excited states of $F_A$(I) (letf) and $F_A$(II) (right) are shown in Figure 4. The latter one plays an essential role in the emission of color center lasers [19], and apparently in some electron dimer $F_A$' centers as well [20]. From Reference [19].

3. Schrödinger equation in 3-D

The eigenvalue vibronic problem is solved by means of $E_{AD-}(Q)$ or $E_{AD+}(Q)$ serving as a potential energy. According;y, there is a vibronic problem in the adiabatic excited state (+) or another one in the adiabatic ground state (-). The lower-energy eigenvalue equation in 3-D reads:

$$-(\hbar^2/2M)\Delta_{Q\theta\varphi} \chi_-(Q) + E_{AD-}(Q)\chi_-(Q) = E_{vib-}\chi_-(Q)$$

$$E_{AD-}(Q) = \tfrac{1}{2}KQ_0^2 - \sqrt{\{(GQ_0)^2 + 2G[(D_c - D_b)\Sigma Q_i^4 + D_b Q_0^4] + \tfrac{1}{4}E_{\alpha\beta}^2\}} =$$

$$-(K/G)[(D_c - D_b)\Sigma Q_i^4 + D_b Q_0^4] - E_{JT}[1 + (E_{\alpha\beta}/4E_{JT})^2] \quad (10)$$

or on introducing small spherical coordinates the vibronic Hamiltonian finally reads

$$H_{vib-3D} = -(\hbar^2/2I)\Delta_{\theta\varphi} - (K/G)Q_0^4[(D_c-D_b)[(\cos\varphi\sin\theta)^4 + (\sin\varphi\sin\theta)^4 + (\cos\theta)^4] + D_b] -$$

$$E_{JT}[1 + (E_{\alpha\beta}/4E_{JT})^2] \quad (11)$$

where $I = MQ_0^2$ is the moment of inertia of the rotating entiity. It may be worth mentioning here that it is not only small-size impurities that displace off-site: Sometimes host ions do likewise, such as the apex oxygens O(A) in high-Tc superconductors of the $La_{1-x}Sr_xCuO_4$ family [12].

The vibronic potential energy has been depicted in Figure 2(a) at pre-calculated values of the entering parameters. We remind that equation (11) refers to the sombrero-hat brim which runs smoothly because of its valley-like character. The rotational barriers are accounted for in Figure 2(b). Apart from the radial derivative $\Delta_r = \partial^2/\partial r^2$, the angular Laplacian $\Delta_{\theta,\varphi}$ plays an essential role in controlling the off-center rotation.

So far, no solution to the (11)-based eigenvalue equation has been reported.

## 4. Schrödinger equation in 2-D

Of particular interest is the off-center rotation in equatorial plane ($\theta = \frac{1}{2}\pi$) which is controlled by the azimuth $\varphi$:

$-(\hbar^2/2I)\,\Delta_\varphi\,\chi_-(\varphi) - (K/G)Q_0^4[(D_c-D_b)[(\cos\varphi)^4 + (\sin\varphi)^4] + D_b]\chi_-(\varphi) =$

$\{E + E_{JT}[1+ (E_{\alpha\beta}/4E_{JT})^2]\}\chi_-(\varphi)$ (12)

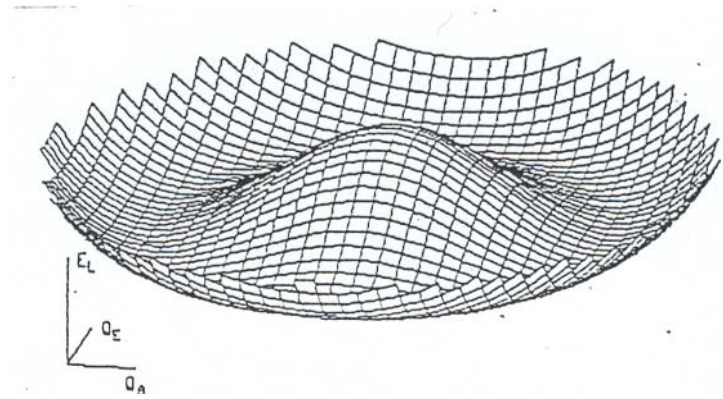

Figure 3: Vibronic potential along the sombrero brim which represents the rotating off-center entity. Chirality will be attached to the orbital rotation along the brim as if to leptons [18].

Following certain manipulations, it has been shown that the resulting eigenvalue equation is Mathieu's equation of a nonlinear rotator. The latter equation corresponds to an in-plane rotation along the brim and may be realized in some solid state systems: $F_A(Li^+)$ center [11].

Using $\cos^4\varphi + \sin^4\varphi = 1 - \frac{1}{2}\sin^2(2\varphi)$, we transform equation (12) to

$-(\hbar^2/2I)\,\Delta_\varphi\,\chi_-(\varphi) + (K/G)Q_0^4[(D_c-D_b)\,\frac{1}{2}\sin^2(2\varphi)] + D_b]\chi_-(\varphi) =$

$\{E + E_{JT}[1+ (E_{\alpha\beta}/4E_{JT})^2] + (1 + D_b)(D_c - D_b)(K/G)Q_0^4\}\chi_-(\varphi)$

We now use $\frac{1}{2}\sin^2(2\varphi) = \frac{1}{4}[1 - \cos(4\varphi)]$ and setting

$q = \frac{1}{4}(K/G)Q_0^4(D_c-D_b)$ (Mathieu's parameter)

we arrive at Mathieu's equation:

$$-(h^2/2I) \Delta_\varphi \chi_-(\varphi) - q \cos(4\varphi) \chi_-(\varphi) =$$

$$\{- q + E + E_{JT}[1 + (E_{\alpha\beta}/4E_{JT})^2] + (1 + D_b)(D_c - D_b)(K/G)Q_0^4\}\chi_-(\varphi) \qquad (13)$$

The conventional form of Mathieu's equation is [13,14]

$$d^2Y/dz^2 + [\lambda + 2q \cos(2z)]Y = 0 \qquad (14)$$

where the relations follow: $q = \pm(2E_{BII} / h\omega_{renII})^{1/2}$, $z = 2\varphi$, $\chi_- = Y$, $\lambda = -E + \ldots$, $2q = a$ [13].

Eigenstates of (13) are the periodic Mathieu functions, while the eigenvalues fall into allowed rotational energy bands [13].

Expanding $\chi_-(\varphi)$ into a power series

$$\chi_-(\varphi) = \exp(2\mu z) \sum_{k=-\infty}^{+\infty} c_k \exp(ikz) \qquad (15)$$

and inserting into equation (13); we get the condition [14]

$$\cosh(\pi\mu) = 1 + 2 \Delta(0) \sin^2(\tfrac{1}{2}\pi\sqrt{\lambda}) \qquad (16)$$

where $\Delta(0)$ is Hill's determinant

$$\Delta(0) = \begin{vmatrix} \ldots & & & & & \ldots \\ \ldots & a/(\lambda-2^2) & 0 & 0 & 0 & \ldots \\ \ldots & 1 & a/(\lambda-1^2) & 0 & 0 & \ldots \\ \ldots & a/\lambda & 1 & a/\lambda & 0 & \ldots \\ \ldots & 0 & a/(\lambda-1^2) & 1 & a/(\lambda-1^2) & \ldots \\ \ldots & 0 & 0 & a/(\lambda-2^2) & 1 & \ldots \\ \ldots & & & & & \ldots \end{vmatrix} \qquad (17)$$

An useful approximation for $\mu$ holds good at small $a = 2q$ which reads

$$\cosh(\pi\mu) \approx 1 + 2\sin^2(\tfrac{1}{2} \pi\sqrt{\lambda}) + [\pi q^2 / (1 - \lambda)\sqrt{\lambda}] \sin(\pi\sqrt{\lambda}) \qquad (18)$$

where $\mu$ is chosen so as to warrant the holding up of (16)- (18).

## 5. Solving for the 3-D eigenvalue equation

There are so far no solutions available to the 3-D problem. This is a problem of the periodic motion of an impurity over the off-center reorientational sites upon the off-center sphere (e.g. Li$^+$ over eight reorientation sites in KCl lattice). The motion is dependent, as we have seen, on the eight interwell barriers separating the metastable reorientation wells. Problems related to the periodic rotation upon a sphere may best be solved by means by Hill's determinant [14].

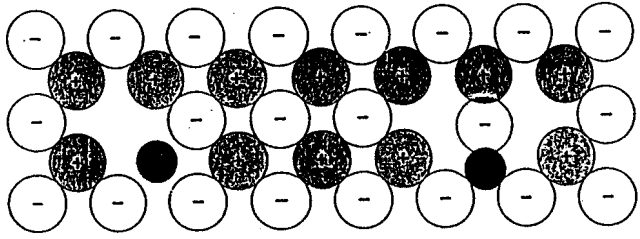

Figure 4: Two alternative configurations at the Li$^+$ impurity in the relaxed excited state of a nearby F center (F$_A$ center) [19].

As a matter of fact, Hill's method [14] may be applied to the 3-D problem, though cautiously, in view of the partial derivatives in (11). The periodic potential of current interest reads ($\cos^4\varphi + \sin^4\varphi = 1 - \tfrac{1}{2}\sin^2(2\varphi)$, $\tfrac{1}{2}\sin^2(2\varphi) = \tfrac{1}{4}[1 - \cos(4\varphi)]$):

$$V(\theta,\varphi) = -(K/G)Q_0^4\{(D_c - D_b)[\,[(\cos\varphi)^4 + (\sin\varphi)^4]\,(\sin\theta)^4 + (\cos\theta)^4] + D_b\}$$

$$= -(K/G)Q_0^4\{(D_c - D_b)[\tfrac{1}{4}[3 - \cos(4\varphi)]\,(\sin\theta)^4 + (\cos\theta)^4] + D_b\} \tag{19}$$

with radial coordinate $Q_0$ and angular coordinates $\theta$ and $\varphi$. Hill's determinant $\Delta(0)$ will be used again leading to

$$\sin^2(\pi k d) = \Delta(0)\sin^2(\pi\sqrt{[2ME']}(d/h) \tag{20}$$

where $d$ is the characteristic constant of the cubic impurity 3-D lattice ingredient.

To first approximation, independent of the specific form of periodic potential,

$$\Delta(0) \approx 1 + 2a^2/\xi^2(1-\xi^2),\ \xi = \sqrt{(2ME')}(d/h),\ a = 2q \tag{21}$$

Following equation (19) we remind that the off-center Hamiltonian is proportional to:

$$H(\theta,\varphi) \propto Q_0^4\,[\tfrac{1}{4}[3 - \cos(4\varphi)]\,(\sin\theta)^4 + (\cos\theta)^4] \tag{22}$$

We will also reproduce a modified form of equations (16) and (20) [15] reading:

$$\sin^2(\tfrac{1}{2}\pi\mu) = \Delta(0)\sin^2(\tfrac{1}{2}\pi\sqrt{\lambda}) \tag{23}$$

The interrelation between the above equations for $\mu$ are easy to disclose.

### 5.1. Evaluating the Z- components

Hamiltonian (22) depends on two angular coordinates. Partially averaging equation (22) over $\theta$ we get

$< H(\theta,\varphi)>|_\theta = Q_0^4 \; ½ \; ¾ \; [¼ \; [3 - \cos(4\varphi)] + 1]$ (24)

This is the familiar Mathieu's nonlinear oscillator Hamiltonian in $\varphi$ close to the formerly obtained result of circular rotation in the equatorial plane ($\theta = ½ \pi$) [11].

Partial averaging over $\varphi$ gives

$< H(\theta,\varphi)>|_\varphi = Q_0^4 \{ ¼ \; [ 3 - \cos(4\theta) ] - ¼ \; (\sin\theta)^4 \} = H_{M\theta} - Q_0^4 \{ ¼ \; (\sin\theta)^4 \}$ (25)

where $H_{M\theta}$ is Mathieu's Hamiltonian along the $\theta$ angular coordinate, while the 2$^{nd}$ term is the Hamiltonian of an anharmonic oscillator $\propto Q^4$.

Schrödinger's equations corresponding to (24) and (25) give:

$< H(\theta,\varphi)>|_\theta \chi_\theta \sim Q_0^4 \; ½ \; ¾ \; \{ ¼ \; [3 - \cos(4\varphi)] + 1 \} \chi_\theta = E_{\theta M} \chi_\theta$ (26)

$< H(\theta,\varphi)>|_\varphi \chi_{\varphi M} \chi_{\varphi A} \sim Q_0^4 \; ¼[3-\cos(4\theta)]\chi_{\varphi M}\chi_{\varphi A} - Q_0^4 \; ¼(\sin\theta)^4 \chi_{\varphi A}\chi_{\varphi M}$

$= E_{\varphi M} E_{\varphi A} \chi_{\varphi M} \chi_{\varphi A}$ (27)

where subscript 'M' stand for Mathieu's and 'A' stand for anharmonic. Mathieu's eigenstates are well-illuminated in the literature.

### 5.2. Approximating for the anharmonic eigenvalues

To first order, the 'A' eigenvalues may be found as

$E_\pm = ½ (H_{11} + H_{22}) \pm ½ \sqrt{[(H_{11} - H_{22})^2 + 4H_{12}H_{21}]}$

where $|1\rangle$ and $|2\rangle$ are appropriate basis states, e.g. displaced harmonic oscillator eigenstates. In this case the matrix elements $H_{ij}$ read $H_{11} \equiv H_{++}$, $H_{22} \equiv H_{--}$, $H_{12} = H_{21} = H_\pm$, to small-polaron approximation:

$H_{++} = \langle 1|H|1 \rangle = A^2 \exp(-\alpha Q_0^2)\sqrt{(\pi/\alpha)}Q_0^4$

$H_{--} = \langle 2|H|2 \rangle = A^2 \exp(-\alpha Q_0^2)\sqrt{(\pi/\alpha)}Q_0^4$

$H_\pm = \langle 2|H|1 \rangle = \langle 1|H|2 \rangle = A^2 \exp(-\alpha Q_0^2)\sqrt{(\pi/\alpha)}Q_0^4$

Accordingly we get

$E_\pm = (K/G)(D_c - D_b)H$

$H = H_{11} \pm H_{12} = (2H_{11}, 0)$

$H_{11} = A^2\sqrt{(\pi/\alpha)}Q_0^4 \exp(-2\alpha Q_0^2)$

$H_{12} = A^2\sqrt{(\pi/\alpha)}Q_0^4 \exp(-2\alpha Q_0^2)$ (27)

### 5.3. Approximating for the anharmonic eigenstate

Using the above equations for the matrix elements we may reproduce the approximate value for the anharmonic wave function $\chi_A = Z_+\chi_+ + Z_-\chi_-$. From the 1$^{st}$ row of the secular determinant we get

$Z_+ = H_{11} - E_- = -H_{12}$

$Z_- = H_{12}$

and, therefore,

$\chi_A(Q) \equiv Z_+\chi_+(Q) + Z_-\chi_-(Q) = -H_{12}\chi_+(Q) + H_{12}\chi_-(Q)$

where $\chi_+(Q) = A\exp(-½\alpha[Q+Q_0])$ and $\chi_-(Q) = A\exp(-½\alpha[Q-Q_0])$ are the basis harmonic-oscillator states, as defined above. It is to be noted that the normalization constant A is still open to manipulation.

### 5.4. Total 3-D eigenstates and eigenvalues

From the above partial $\varphi$-averaging analysis along the $\theta$-coordinate we learn that the total 3-D eigenstate for the Li$^+$ off-center problem is composed of a Mathieu eigenfunction $Z(\theta)$ times an anharmonic $Z(Q^4)$ eigenstate acting along the off-center ring. Accordingly, the eigenvalues are composed of Mathieu's periodic eigenenergies in allowed rotational bands. These are complemented by anharmonic eigenvalues which have only been derived approximately to 1$^{st}$ order.

Our analysis has further shown that the 2-D eigenproblem for rotation along the off-center ring that has formerly been sought as rotation in the equatorial plane needs minor corrections only, now under a partial averaging over the $\theta$- coordinate.

Both analyses are to be refined by more stringent calculations which go beyond the scope of the present study.